\documentclass[epj-spec,nopacs]{svjour}
\usepackage{graphics}
\usepackage{amsmath,amssymb,graphics}

\begin{document}

\title{Physical States and BRST Operators for Higher-spin $W$ Strings}
\author{Yu-Xiao Liu\inst{1}\fnmsep\thanks{\email{liuyx@lzu.edu.cn}} \and
        Shao-Wen Wei \inst{1}\fnmsep\thanks{Corresponding author. \email{weishaow06@lzu.cn}} \and
        Li-Jie Zhang \inst{2}\fnmsep\thanks{\email{lijzhang@shu.edu.cn}} \and
        Ji-Rong Ren \inst{1}}
\institute{
    Institute of Theoretical Physics, Lanzhou University,
           Lanzhou 730000, P. R. China \and
    Department of Physics, Shanghai University,
           Shanghai 200444, P. R. China}

\abstract{ In this paper, we mainly investigate the
$W_{2,s}^{M}\otimes W_{2,s}^{L}$ system, in which the matter and
the Liouville subsystems generate $W_{2,s}^{M}$ and $W_{2,s}^L$
algebras respectively. We first give a brief discussion of the
physical states for corresponding $W$ stings. The lower states are
given by freezing the spin-2 and spin-$s$ currents. Then,
introducing two pairs of ghost-like fields, we give the
realizations of $W_{1,2,s}$ algebras. Based on these linear
realizations, BRST operators for $W_{2,s}$ algebras are obtained.
Finally, we construct new BRST charges of Liouville system for
$W_{2,s}^{L}$ strings at the specific values of central charges
$c$: $c=-\frac{22}{5}$ for $W_{2,3}^{L}$ algebra, $c=-24$ for
$W_{2,4}^{L}$ algebra and $c=-2,-\frac{286}{3}$ for $W_{2,6}^{L}$
algebra, at which the corresponding $W_{2,s}^L$ algebras are
singular.}



\maketitle

\section{Introduction}
\label{secIntroduction}

To well understand the properties of a string theory, one needs to
obtain its underlying word-sheet symmetry algebra. Since $W$
algebras \cite{Zamolodchikov1985,FateevNPB1987} received
considerable attention and application, much work
\cite{BoerCMP1994,DeckmynPRD1995,BoreschNPB1995,MadsenCMP1997,Pope1991,BergshoeffPLB1990,BergshoeffNPB1994,PopePLB1991}
has been carried out on the classification of $W$ algebras and the
study of $W$ gravities and $W$ strings. In fact, $W$ algebras also
imply some underlying symmetry and appear in the quantum Hall
effect \cite{KarabaliNPB1994} and black holes
\cite{IsoPRD2007,BonoraJHEP2008}, in lattice models of statistical
mechanics at criticality, and in other physical models
\cite{Leznov1989,FeherPR1992} and so on.

In all applications of $W$ algebras, the investigation of $W$
strings is more interesting and important. The idea of building
$W$ string theories was first developed in Ref.
\cite{GervaisNPB1987}. Since then, much research on the scalar
realization of $W_{2,s}$ strings has been done
\cite{ThierryPLB1987,BaisNPB1988,SchransNPB1993,LuCQG1994,Lu2IJMPA1994,Lu3CQG1994,BershadskyPLB1992,BergshoeffPLB1993}.
In \cite{PopePLB1993,Pope1993}, the states for $W_{2,3}$ had been
studied. The spectrum of $W$ strings also can be found in
\cite{DasMPLA1991,PopePLB1992,LuNPB1995,LuIJMP1993,LuNPB1993,WestIJMPA1993}.
The spinor realizations of $W_{2,s}$ strings were given in
\cite{ZhaoPLB2000,DuanNPB2004,LiuJHEP2005,ZhangCTP2006}. It is
known that, when extended to the quantum case, $W_{2,s}$ algebras
will become non-linear. Fortunately, some of these algebras could
be linearized by the inclusion of a spin-1 current. For the cases
of $s=3$ and $4$, the linearizations were obtained in
\cite{LiuJHEP2005,ZhangCTP2006,LuIIMPLA1995,KrivonosPLB1994,LuMPLA,Madsen1995}.
There exists no linear $W_{1,2,5}$ algebra at the quantum level.
However $W_{2,6}$ algebra can be linearized as $W_{1,2,6}$ algebra
when the central charge takes the specific value $c=390$, and a
realization for $W_{1,2,6}$ algebra was given in
\cite{LuIIMPLA1995}. In this paper, by introducing a pair of
bosonic ghost-like fields $(R, S)$ with spins $(6,-5)$ and a pair
of fermionic ghost-like fields $(b_{1},c_{1})$ with spins
$(k,1-k)$, we will investigate the realization and the BRST
operator of $W_{2,6}$ algebra at the central charge $c=390$.

Noncritical strings are strings in which the two-dimensional
gravitational fields do not decouple after quantization but
instead develop an induced kinetic term. The string coordinates
are called `matter' fields while the non-decoupled gravitational
fields are represented by a set of so-called `Liouville' fields.
In \cite{BergshoeffNPB1994,BergshoeffPLB1995}, the BRST operators
for the Liouville system were obtained. Especially, in
\cite{LuIIMPLA1995,LuIPLB1995}, the BRST operators of
$W_{2,s}^{M}\otimes W_{2,s'}^{L}$ were constructed at the
classical level, and much valuable results were given. In general,
classical $W_{2,s}$ algebras are regular. But when extended to the
quantum case, they will become singular at some values of the
central charge, for example, $c=-\frac{22}{5}$ for $W_{2,3}$
algebra, $c=-24$, $\frac{1}{2}$, $-\frac{22}{5}$, $-\frac{68}{7}$
for $W_{2,4}$ algebra, and $c=-2$, $-\frac{286}{3}$ for $W_{2,6}$
algebra. But one can rescale the spin-3, spin-4 and spin-6
currents such that the corresponding OPEs have no divergent
coefficients. It was proved that at $c=-\frac{22}{5}$ for
$W_{2,3}$, $c=-24$ for $W_{2,4}$, and $c=-2$, $-\frac{286}{3}$ for
$W_{2,6}$, these algebras satisfy the Jacobi identity
\cite{LuIPLB1995}. This implies that we can construct the explicit
BRST operators for $W_{2,3}^{L}$, $W_{2,4}^{L}$ and $W_{2,6}^{L}$
strings at $c_{L}=-\frac{22}{5}$, $c_L=-24$ and $c_L=-2$,
$-\frac{286}{3}$ respectively. In this paper, we will construct
the BRST operators for $W_{2,s}^{L}$ string by using grading
method.

The present paper is organized as follows. In section
\ref{Physical state}, we give a brief discussion of the physical
state for the $W_{2,s}^{M}\otimes W_{2,s}^{L}$ system. Then in
section \ref{BRST operators}, we build the BRST operators for
$W_{2,s}$ algebras by giving the realizations for the linear
$W_{1,2,s}$ algebras. In section \ref{BRST operator}, using the
grading method, we obtain the BRST operators for Liouville system
of $W_{2,s}^{M}\otimes W_{2,s}^{L}$ for $s=3$, $4$, and $6$. And
finally, the paper ends with a brief conclusion.

\section{Physical states for the $W_{2,s}^{M}\otimes W_{2,s}^{L}$ system}
\label{Physical state}

The physical states in string theory can most elegantly be
described by using the BRST formalism. For bosonic string, the
BRST operator $Q_B$ is given by
\begin{eqnarray}
Q_B= \oint dz c(z)\bigg(T(z)+\frac{1}{2} T^{gh}(z) \bigg).
\end{eqnarray}
There also exists a similar expression for anti-holomorphic part.
Here $T(z)$ is the energy-momentum tensor for the matter, and
$T^{gh}$ the ghost energy-momentum tensor with central charge
$-26$.

However for $W_{2,s}^{M}\otimes W_{2,s}^{L}$ algebra, which has
two series systems, i.e., the matter and the Liouville systems,
the BRST operator for this case is given by
\begin{eqnarray}
 Q_{B}&=&Q_{B}^{M}+Q_{B}^{L}, \\
 Q_{B}^{M}&=&\oint dz \left[c^{M}(z)T^{M}(z)
             +\gamma ^{M}(z) W^{M}(z)\right], \\
 Q_{B}^{L}&=&\oint dz \left[c^{L}(z)T^{L}(z)
             +\gamma ^{L}(z)W^{L}(z)\right],\label{QL}
\end{eqnarray}
where $T^{M}$ and $W^{M}$ are the currents of matter system for
$W_{2,s}^M$ algebras and have spin-2 and spin-$s$ respectively,
while $T^{L}$ and $W^{L}$ are the currents of Liouville system for
$W_{2,s}^L$ algebras. Note that the ghost sector is included in
$T^{M}$ and $W^{M}$, and also in $T^{L}$ and $W^{L}$. The
fermionic ghosts $(b,c)$ and $(\beta,\gamma)$ are introduced for
the currents $T$ and $W$. The up index `$M$' denotes matter and
`$L$' Liouville. In fact, we ignore the anti-holomorphic sector
and just focus on the holomorphic one.

Physical states are defined to be states in the cohomology of BRST
operator $Q_{B}$. The entire linear space $H=H^{M}{\oplus}H^{L}$
of the operator $Q_{B}=Q_{B}^{M}+Q_{B}^{L}$ can also be
decomposed, with respect to a grading naturally associated with
the underlying affine algebra, as $H = H_{-}\oplus H_{+}$
\cite{GrassiJHEP2002,KacFAA1967}. $H_{-}$ and $H_{+}$ have
negative and non-negative grading, respectively. The matter
subspace $H^M$ and Liouville subspace $H^L$ can also be decomposed
as $H^M = H_{-}^M\oplus H_{+}^M$ and $H^L= H_{-}^L\oplus H_{+}^L$
with $H^{M}_{+}$ and $H^{L}_{+}$ the matter and Liouville
subspaces with non-negative gradings. Then the physical state
condition is expressed as
\begin{equation}
Q^{M}_{B} |\psi \rangle =Q^{L}_{B} |\psi \rangle =0, \quad
|\psi\rangle \in H^{M}_{+} \otimes H^{L}_{+},
\end{equation}
and
\begin{equation}
\begin{array}{l}
|\psi \rangle \neq Q^{M}_{B} |\varphi \rangle +Q^{L}_{B} |\phi
\rangle, \quad |\varphi \rangle, |\phi \rangle \in H^{M}_{+}
\otimes H^{L}_{+}, \\
Q^{M}_{B} |\phi \rangle =Q^{L}_{B} |\varphi \rangle=0.
\end{array} \label{state condition}
\end{equation}
In this system, physical states have the form
\begin{equation}
|\psi \rangle = |\texttt{phys} \rangle_{M} \otimes |\downarrow
\rangle_{M} +|\texttt{phys} \rangle_{L} \otimes |\downarrow
\rangle_{L},\label{standard}
\end{equation}
where $|\texttt{phys} \rangle_{M}$ involves operators built from
the matter system. The matter system ghost vacuum $| \downarrow
\rangle_{M}$ is built from ghost fields $(b^{M},c^{M})$ and
$(\beta^{M},\gamma^{M})$. It is similar to the Liouville system.

For the latter, we just consider the Liouville system and ignore
the matter one which is equivalent to the Liouville system. Here,
we write down these currents in Laurent modes
\begin{eqnarray}
 T^{L}(z)&=&\sum_{n=-\infty}^{+\infty}L_{n}^{L} z^{-n-2},
            \label{Laurent1} \\
 W^{L}(z)&=&\sum_{n=-\infty}^{+\infty}W_{n}^{L} z^{-n-s},
\end{eqnarray}
where $s$ is the conformal spin of the current $W^{L}$.
The modes $L_{n}^{L}$ generate the Virasoro algebra:
\begin{equation}
[L_{m},L_{n}]=(m-n)L_{m+n}+\frac{c}{12}m(m^{2}-1)\delta_{m+n,0},
\label{LD}
\end{equation}
where $c$ is the central charge. For critical strings, by
introducing the ghost fields $(b,c)$ for the spin-2 current, the
central charge $c$ will vanish and (\ref{LD}) will reduce to the
classical case.

The ghost fields $(b,c)$ and $(\beta,\gamma)$ in Laurent modes are
\begin{eqnarray}
 b^{L}(z)&=&\sum_{n=-\infty}^{+\infty}b_{n}^{L} z^{-n-2}, ~\quad
 c^{L}(z)=\sum_{n=-\infty}^{+\infty}c_{n}^{L} z^{-n+1},
         \label{Laurent1.5}\\
 \beta^{L}(z)&=&\sum_{n=-\infty}^{+\infty}\beta_{n}^{L} z^{-n-s},\quad
 \gamma^{L}(z)=\sum_{n=-\infty}^{+\infty}\gamma_{n}^{L}z^{-n-(1-s)}.
          \label{Laurent2}
\end{eqnarray}
It is easy to see that the conformal spin of $(b,c)$ and
$(\beta,\gamma)$ are $(2,-1)$ and $(s,1-s)$ respectively.
The ghost vacuum $|\downarrow \rangle_{L}$ is given by
\begin{equation}
|\downarrow \rangle_{L} = c^{L}_{1}
\gamma^{L}_{1}\gamma^{L}_{2}\cdots \gamma_{s-1}^{L}|0\rangle,
\end{equation}
which is obtained from the $SL(2,C)$-invariant vacuum $|0\rangle$,
which satisfies
\begin{eqnarray}
b_{n}^{L}|0\rangle&=&0,\;\;\; n \geq -1, \quad\quad\;\;\; c_{n}^{L}|0\rangle=0,\;\;\;n \geq 2, \\
\beta_{n}^{L}|0\rangle&=&0,\;\;\;n \geq -s+1, \quad
\gamma_{n}^{L}|0\rangle=0,\;\;\;n \geq s .
\end{eqnarray}
For the states of the form $(\ref{standard})$, the condition of
BRST invariance becomes
\begin{eqnarray}
L_{0}^L|\texttt{phys}\rangle &=&\Delta |\texttt{phys}\rangle, \\
W_{0}^L|\texttt{phys}\rangle &=&\sigma_{s} |\texttt{phys}\rangle,\\
L_{n}^L|\texttt{phys}\rangle &=&
W_{n}^L|\texttt{phys}\rangle=0,\quad n\geq 1,
\end{eqnarray}
where the constants $\Delta$ and $\sigma_{s}$ are the intercepts
for the zero modes of the spin-2 and spin-$s$ currents
respectively. For $W_{2,4}$ string $\Delta=4$, and for bosonic
string $\Delta=1$. The intercepts $\sigma_{s}$ of the spin-$s$
currents are given in \cite{LuNPB1995}.

Using Eqs. (\ref{Laurent1})-(\ref{Laurent2}), the BRST operator
can be written in Laurent modes
\begin{equation}
Q_{B}^{L}=\sum_{n=-\infty}^{+\infty}
(c^L_{n}L_{-n}^L+\gamma^L_{n}W^L_{-n}). \label{QB}
\end{equation}
We denote the physical ground as $|\downarrow,k\rangle$ which is
characterized by level number $\ell=0$ and ghost number $G=0$.
Then one will get excited states by the action of descendent
operators like $c^L_{-n}, b^L_{-n}, \beta^L_{-n}, \gamma^L_{-n}$,
$L^L_{-n}$ and $W^L_{-n}$ $(n>0)$ on $|\downarrow,k\rangle$. In
fact, in order to get the physical states, the known of detail
$W_{2,s}^L$ algebra is necessary. But we can `freeze' the spin-2
current $T^L$ and spin-$s$ current $W^L$, which means no creation
operators from them can appear in physical states. Then the
physical states up to level $\ell= 2$ are as follows:
\begin{eqnarray}
 \ell&=&0: \quad |\downarrow,k\rangle, \nonumber \\
 \ell&=&1: \quad (p_{1}c^L_{-1}+p_{2}\gamma^L_{-1})|\downarrow,k\rangle,
           \label{state} \\
 \ell&=&2: \quad (p_{3}c^L_{-2}+p_{4}\gamma^L_{-2}
         +p_{5}c^L_{-1}\gamma^L_{-1})
         |\downarrow,k\rangle, \nonumber
\end{eqnarray}
where $p_{1}-p_{5}$ are constants.

Now, we would like to end this section with some comments. It is
worth to note that the sum of down-index in each term in
(\ref{QB}) is zero. This lead to a special property, i.e., the
level of an arbitrary physical state does not change when acting
it with $Q_{B}$. The physical states we given above are based on
the condition of `freezing' the spin-2 and spin-$s$ currents. The
more detail of the physical states needs the explicit OPEs of
$W_{2,s}^L$ algebra.

\section{BRST operators for $W_{2,s}$ algebras from $W_{1,2,s}$ algebras}
\label{BRST operators}

After a brief investigation of the physical states, we would like
to construct BRST operators for $W_{2,s}$ algebras from the linear
$W_{1,2,s}$ algebras.

\subsection{BRST operators for $W_{2,3}$ and $W_{2,4}$ algebras}

At the classical level, $W_{2,s}$ algebras exist for all positive
integer values of $s$. These algebras are generated by the spin-2
energy-momentum tensor $T$ and a primary spin-$s$ current $W$,
which satisfy the OPEs
\begin{eqnarray}
 T(z) T(\omega)~ &\sim& \frac{2T}{(z-\omega)^{2}}
       +\frac{\partial T}{z-\omega}, \nonumber \\
 T(z) W(\omega) &\sim&  \frac{s W}{(z-\omega)^{2}}
       +\frac{\partial W}{z-\omega},\\
 W(z) W(\omega) &\sim& \frac{2T^{s-1}}{(z-\omega)^{2}}
       +\frac{\partial T^{s-1}}{z-\omega}.\nonumber
\end{eqnarray}
The corresponding BRST operator is given by \cite{LuIPLB1995}
\begin{equation}
Q_{B}=\oint dz \big[c \big(T-s \beta \partial \gamma
-(s-1)\partial \beta \gamma -b\partial c\big)+\gamma W-\partial
\gamma \gamma b T^{s-2}\big], \label{QBclassical}
\end{equation}
where $(b,c)$ and $(\beta,\gamma)$ are the ghost and anti-ghost
fields for the currents $T$ and $W$ respectively.

When extending these algebras to the quantum case, the BRST
operator $Q_{B}$ will not be the form of (\ref{QBclassical}) since
these algebras are not linear anymore. In fact the OPE of two
currents with spins $s$ and $s'$ produces terms, at leading order,
with spin $s+s'-2$. For example, there will be terms with spin 4
in the OPEs of $W_{2,3}$ algebra. But these terms with spin
$s+s'-2$ can be interpreted as composite fields built from the
products of the fundamental currents with spin $s$ and $s'$. The
OPE $W(z)W(\omega)$ for $W_{2,3}$ algebra is given by
\begin{equation} \aligned
 W(z)W(\omega) \sim & \frac{c/3}{(z -w)^6}
   +\frac{2T}{(z-\omega)^4}+ \frac{\partial T}{(z-\omega)^3}\\
   +&\frac{1}{(z-\omega)^2}\left(2 \Theta \Lambda
     +\frac{3}{10}\partial^2 T \right)
     +\frac{1}{(z-\omega)}\left(\Theta \partial \Lambda
     +\frac{1}{15}\partial ^3 T \right),
\endaligned \end{equation}
where
\begin{equation} \aligned
 \Theta= \frac{16}{22+5c},\;\;\;\;\;\; \Lambda = T^2
- \frac{3}{10} \partial ^2 T.
\endaligned \end{equation}
For the case $W_{2,4}$, the OPE $W(z)W(\omega)$ takes the form
\begin{equation}
 \aligned
 W(z)W(\omega) \sim & \bigg\{\frac{2 T}{(z - w)^6}
   +\frac{\partial T}{(z-\omega)^5}
   +\frac{3}{10} \frac{\partial ^2 T}{(z-\omega)^4}\\
 + & \frac{1}{15}\frac{\partial ^3 T}{(z-\omega)^3}
   +\frac{1}{84}\frac{\partial ^4 T}{(z-\omega)^2}
   +\frac{1}{560}\frac{\partial ^5 T}{(z-\omega)}\bigg\}\\
 + & \sigma_{1} \bigg\{\frac{U}{(z-\omega)^4}
   +\frac{1}{2} \frac{\partial U}{(z-\omega)^3}
   + \frac{5}{36} \frac{\partial ^2 U}{(z-\omega)^2}
   +\frac{1}{36} \frac{\partial ^3 U}{(z-\omega)} \bigg\}\\
 + & \sigma_{2} \bigg\{ \frac{W}{(z-\omega)^4}
   +\frac{1}{2}\frac{\partial W}{(z-\omega)^3}
   + \frac{5}{36} \frac{\partial ^2 W}{(z-\omega)^2}
   +\frac{1}{36} \frac{\partial ^3 W}{(z-\omega)} \bigg\}\\
 +& \sigma_{3} \bigg\{ \frac{G}{(z-\omega)^2}
   +\frac{1}{2}\frac{\partial G}{(z-\omega)} \bigg\}
   + \sigma_{4}\bigg\{ \frac{A}{(z-\omega)^2}
   +\frac{1}{2}\frac{\partial A}{(z-\omega)}\bigg\}\\
 + & \sigma_{5}\bigg\{ \frac{B}{(z-\omega)^2}
   +\frac{1}{2} \frac{\partial B}{(z-\omega)}\bigg\}
   +\frac{c/4}{(z-\omega)^8},
 \endaligned
\end{equation}

\noindent where the composites $U$ (spin 4), and $G$, $A$ and $B$
(all spin 6), are defined by
\begin{equation}
\aligned
 U&=(TT)-\frac{3}{10} \partial ^2 T,\;\;\;\;\;\; G=(\partial ^2 T
T)-\partial (\partial T T)+\frac{2}{9} \partial ^2
(TT)-\frac{1}{42}\partial ^4 T,\\
A & =(T U)-\frac{1}{6}\partial ^2 U, \;\;\;\;\;\; B=(T
W)-\frac{1}{6}\partial ^2 W ,
\endaligned
\end{equation}
with normal ordering of products of currents understood. The
coefficients $\sigma_{i}(i=1-5)$ are given by
\begin{equation}
\aligned
 \sigma_{1} & =\frac{42}{5c+22},\;\;\;\;\;\;
\sigma_{2}=\sqrt{\frac{54(c+24)(c^2 -172c
+196)}{(5c+22)(7c+68)(2c-1)}},\\
\sigma_{3} & =\frac{3(19c-524)}{10(7c+68)(2c-1)},\;\;\;\;\;\;
\sigma_{4}
=\frac{24(72c+13)}{(5c+22)(7c+68)(2c-1)},\\
\sigma_{5} & = \frac{28}{3(c+24)} \sigma_{2}.
\endaligned
\end{equation}
It is worth to point out that, at the quantum level, $W_{2,3}$
algebra with central charge $c=-\frac{22}{5}$ and $W_{2,4}$
algebra with $c=-24$, $\frac{1}{2}$, $-\frac{22}{5}$ or
$-\frac{68}{7}$, are singular.

By introducing a spin-1 current $J_0$, nonlinear $W_{2,3}$ and
$W_{2,4}$ algebras can be linearized as $W_{1,2,3}$ and
$W_{1,2,4}$ algebras, respectively. The linear $W_{1,2,s}$
algebras for $s=3,4$ take the form \cite{KrivonosPLB1994}
\begin{eqnarray}
 T_{0}(z) T_{0}(\omega) &\sim& \frac{c/2}{(z-\omega)^{4}}
   +\frac{2T}{(z-\omega)^{2}}+\frac{\partial T}{z-\omega}, \quad
 T_{0}(z) W_{0}(\omega) \sim  \frac{s W}{(z-\omega)^{2}}
   +\frac{\partial W}{z-\omega},\nonumber\\
 T_{0}(z)J_{0}(\omega) &\sim& \frac{c1}{(z-\omega)^{3}}
   +\frac{J_{0}}{(z-\omega)^{2}}+\frac{\partial J_{0}}{z-\omega},
   \quad
 J_{0}(z)J_{0}(\omega) \sim -\frac{1}{(z-\omega)^{2}}, \label{OPE} \\
 J_{0}(z)W_{0}(\omega) &\sim& \frac{h W_{0}}{z-\omega},\quad
 W_{0}(z)W_{0}(\omega) \sim 0. \nonumber
\end{eqnarray}
From these OPEs, it is clear that the current $W_{0}$ is a primary
field, but $J_{0}$ not. The coefficients $c$, $c1$ and $h$ are
given by
\begin{eqnarray}
\aligned
 &c=50+24t^{2}+\frac{24}{t^{2}},\quad
  c1=-\sqrt{6}(t+\frac{1}{t}),\quad
  h=\sqrt{\frac{3}{2}}t,\quad(s=3)\\
 &c=86+30t^{2}+\frac{60}{t^{2}},\quad
  c1=-3t-\frac{4}{t},\quad \quad\;\;
  h=t.\quad\quad\quad(s=4)
\endaligned
\end{eqnarray}

The bases $T$ and $W$ of $W_{2,s}$ algebras are constructed by the
linear bases of the $W_{1,2,s}$ algebras in our previous paper
\cite{LiuJHEP2005}:
\begin{eqnarray}
 T &=&T_0 , \label{W2324T} \\
 W &=&W_{0}+\zeta_1 \partial^2 J_{0}+\zeta_2 \partial J_{0}J_{0}
     +\zeta_3 J_{0}^{3}+\zeta_4 \partial T_{0}
     +\zeta_5 T_{0}J_{0}, \quad\quad\quad\quad ~~~(s=3)  \label{W23X} \\
 W &=& W_0 + \eta_1 \partial^3 J_0 + \eta_2 \partial^2 J_0J_0
     +\eta_3(\partial J_0)^2 + \eta_4 \partial J_0(J_0)^2
     +\eta_5(J_0)^4   \nonumber \\
   &&+\eta_6 \partial ^2 T_0 + \eta_7(T_0)^2
     + \eta_8 \partial T_0 J_0 +\eta_9T_0 \partial J_0
     + \eta_{10} T_0(J_0)^2, \quad\quad\quad (s=4) \label{W24X}
\end{eqnarray}
where the realization of $T_0$, $J_0$ and $W_0$ as well as the
coefficients $\zeta_i$ and $\eta_i$ can be found in
\cite{LiuJHEP2005}. Note that this linearization does not contain
the case $c=-\frac{22}{5}$ for $W_{2,3}$, and the cases $c=-24$,
$\frac{1}{2}$, $-\frac{22}{5}$ and $-\frac{68}{7}$ for $W_{2,4}$,
for which the algebra are singular. After careful calculation, we
find that there exists no linearization for these singular
algebras. However, for these specific values of central charge, we
can rescale the spin-3 and spin-4 currents such that the OPEs have
no divergent coefficients. This will be discussed in detail in
next section.

The BRST operator for a $W_{2,s}$ algebra is given by
\begin{equation} \label{QBW2sAlgebras}
Q_{B}=\oint dz [c(z) T(z)+\gamma(z) W(z) ],
\end{equation}
Substituting (\ref{W2324T})-(\ref{W24X}) into
(\ref{QBW2sAlgebras}), we will obtain the explicit forms of the
BRST operators for $W_{2,3}$ and $W_{2,4}$ algebras.

\subsection{BRST operator for $W_{2,6}$ algebra}
 For the higher spin case of $W_{2,5}$, there is no such
linearization as $W_{2,3}$ and $W_{2,4}$. However, the spin-6
current $W$ can be linearized as \cite{LuIIMPLA1995}
\begin{equation}\aligned
 W =& W_{0}-\frac{1}{6} J_{0}^{6}-\frac{1}{2}T_{0}J_{0}^{4}-\frac{4921}{114718}T_{0}^{3}
            -\frac{3}{8}T_{0}^{2}J_{0}^{2}+\frac{9}{8}T_{0}^{2}\partial J_{0}
            +\frac{15}{2}T_{0}\partial J_{0}J_{0}^{2}-\frac{21}{2}T_{0}(\partial J_{0})^{2} \\
            &-\frac{41}{4}T_{0} \partial^{2}
            J_{0}J_{0}+\frac{21}{4}T_{0}\partial^{3}J_{0}+\frac{11}{2}\partial
            J_{0} J_{0}^{4}-\frac{315}{8} (\partial J_{0})^{2}
            J_{0}^{2}+\frac{277}{8}(\partial J_{0})^{3}+\frac{7}{4}
            \partial T_{0}J_{0}^{3}\\
            &+\frac{3}{2} \partial T_{0}T_{0}J_{0}-\frac{57}{4}
            \partial T_{0}\partial
            J_{0}J_{0}-\frac{190257}{229436}(\partial
            T_{0})^{2}+\frac{43}{4}\partial T_{0} \partial^{2}
            J_{0}-\frac{157}{12}\partial^{2}J_{0}J_{0}^{3}\\
            &+\frac{409}{4}\partial^{2} J_{0} \partial
            J_{0}J_{0}-\frac{1763}{48}(\partial^{2}J_{0})^{2}
            -\frac{108753}{114718}\partial^{2}T_{0}T_{0}-\frac{45}{16}\partial^{2}T_{0}J_{0}^{2}
            +\frac{135}{16}\partial^{2}T_{0}\partial J_{0}\\
            &+\frac{273}{16}\partial^{3}J_{0}J_{0}^{2}-\frac{787}{16}\partial^{3}J_{0}\partial J_{0}
            +\frac{5}{2}\partial^{3}T_{0}
            J_{0}-\frac{197}{16}\partial^{4}
            J_{0}J_{0}-\frac{440915}{458872}\partial^{4}T_{0}+\frac{383}{96}\partial^{5}J_{0}.
            \label{W6}
\endaligned \end{equation}
The currents $T_{0}$, $J_{0}$, and $W_{0}$ have spin 2, 1, and 6
respectively. They generate the linear $W_{1,2,6}$ algebra which
also takes the form of (\ref{OPE}), but the coefficients are given
by
\begin{equation}
c=390,\quad c1=11,\quad h=-1,\quad s=6.
\end{equation}
Note that this is different from the cases of $W_{2,3}$ and
$W_{2,4}$, for these constants take specific values.

To obtain a new realization for the linear $W_{1,2,6}$ algebra, we
introduce a pair of bosonic ghost-like fields $(R, S)$ with spins
$(6,-5)$ and a pair of fermionic ghost-like fields $(b_{1},c_{1})$
with spins $(k,1-k)$ to construct the linear bases of it. The
realization for the $W_{1,2,6}$ algebra is given by
\begin{eqnarray}
 T_{0}&=&T_{eff}-T_{g},\nonumber \\
 J_{0}&=&\rho\; RS+\lambda \; b_{1}c_{1}, \\
 W_{0}&=&R,\nonumber
\end{eqnarray}
where
\begin{equation}
T_{g}=6\;R\partial S+5\;\partial R S+k\; b_{1}\partial
c_{1}+(k-1)\;\partial b_{1}c_{1}, \label{T0g}
\end{equation}
and $T_{g}$ and $T_{eff}$ have central charges $c_{g}$ and
$c_{eff}$, respectively. By making use of the OPEs
$J_{0}(z)J_{0}(\omega)$ and $J_{0}(z)W_{0}(\omega)$ in
(\ref{OPE}), we can solve the coefficients $\rho$ and $\lambda$.
And the value of $k$ is determined from the OPE relation of
$T_{0}$ and $J_{0}$. Substituting this value into (\ref{T0g}), we
get the value of $c_{g}$. Since the central charges $c_{g}$ and
$c_{eff}$ satisfy the condition $c_{eff}+ c_{g}=390$, the value of
$c_{eff}$ can be obtained. All the coefficients are listed as
follows:
\begin{equation}
k=1,\;\;\lambda=0,\;\;\rho=-1,\;\;c_{g}=360,\;\;c_{eff}=30.
\label{W0}
\end{equation}
Substituting (\ref{W0}) into (\ref{W6}), one can obtain the spin-6
current $W$, which together with the $T=T_{0}$ generate the
$W_{2,6}$ algebra. Substituting $T$ and $W$ into
(\ref{QBW2sAlgebras}), we will obtain the explicit form of the
BRST operator $Q_{B}$ for $W_{2,6}$ algebra.

\section{BRST operators of Liouville system for $W_{2,s}^{L}$ strings}
\label{BRST operator}

As shown in section \ref{BRST operators}, at $c_{L}=-\frac{22}{5}$
for $W_{2,3}^{L}$ algebra and $c_{L}=-24$, $\frac{1}{2}$,
$-\frac{22}{5}$, $-\frac{68}{7}$ for $W_{2,4}^L$ algebra, these
algebras will become singular. But one can rescale the spin-3 and
spin-4 currents such that their OPEs have no divergent
coefficients \cite{LuIPLB1995}. One can prove that at
$c_{L}=-\frac{22}{5}$ for $W_{2,3}^L$ and $c_{L}=-24$ for
$W_{2,4}^L$, these algebras satisfy the Jacobi identity. In this
section, we will give the explicit BRST operators for the
corresponding $W_{2,s}^{L}$ strings.

Now, we turn our attention to the grading method. The BRST
operator can be rewritten in the grading form
\begin{eqnarray}
 Q_{B}&=&Q_{0}+Q_{1}, \label{gradingQB}\\
 Q_{0}&=&\oint dz c {\cal T}(\phi,b,c,\beta,\gamma,T_L), \\
 Q_{1}&=&\oint dz \gamma {\cal W}(\phi,\beta,\gamma,T_L,W_L), \label{Wconstruct}
\end{eqnarray}
where we introduce the $(b, c)$ ghost system for the spin-2
current $T_L$, and the $(\beta,\gamma)$\ ghost system for the
spin-$s$ current $W_L$. The ghost fields $b$, $c$, $\beta$,
$\gamma$ are all fermionic and anticommuting. They satisfy the
OPEs
\begin{equation}
 b(z) c(\omega) \sim \frac{1}{z-\omega}, \quad \beta(z) \gamma(\omega) \sim
    \frac{1}{z-\omega}.
\end{equation}
In other cases the OPEs vanish. The nilpotency conditions are
\begin{equation}
Q_{0}^{2}=Q_{1}^{2}=\{ Q_{0},Q_{1} \}=0. \label{nicon}
\end{equation}
One could see that, with this grading method, the construction of
BRST operators will become easy. However, it imposes some
restrained conditions on the BRST operators.

Next, we would like to construct the explicit BRST operators of
Liouville system for $W_{2,s}$ strings for $s=3,4,6$ by using the
grading method. For simplify, we ignore the index `$L$' in $b^{L},
c^{L}, \beta^{L}$ and $\gamma^{L}$ which denotes the Liouville
system.

\subsection{BRST operator for $W_{2,3}^{L}$ string}

At $c_{L}=-\frac{22}{5}$, the quantum $W_{2,3}^L$ algebra will
become singular. But after rescaling the spin-3 current, the OPE
of $W_L(z)W_L(\omega)$ reads \cite{LuIPLB1995}
\begin{equation}
 W_L(z)W_L(\omega) \sim \frac{2(T_L^{2}-\frac{3}{10}
          \partial^{2}T_L)}{(z-\omega)^{2}}
          +\frac{\partial T_L^{2}-\frac{3}{10}
          \partial^{3} T_L}{z-\omega}.
\end{equation}
It can be verified that $T_L$ and $W_L$ define a consistent
algebra which satisfies the Jacobi identity. This implies that it
can be used to construct the non-critical BRST operator for
$W_{2,3}^{M}\otimes W_{2,3}^{L}$ at $c_{L}$=$-\frac{22}{5}$.

In Ref. \cite{wei2008}, we obtained four solutions for the first
grading BRST operator $Q_{0}$. Here we choose the following
solution:
\begin{equation}
 Q_{0}=\oint dz c\bigg(T_{eff}+T_{L}
     +T_{\phi}+\frac{1}{2}T_{bc}
     +T_{\beta\gamma}\bigg), \label{W23Q0}
\end{equation}
where $T_{eff}$ is an effective energy-momentum tensor with
central charge $c_{eff}$. The other energy-momentum tensors are
given by
\begin{eqnarray}
T_{\phi}&=&-\frac{1}{2}(\partial \phi)^{2}-q
\partial^{2}\phi, \\
T_{bc}&=&-2b\partial c-\partial bc, \label{TBG} \\
T_{\beta\gamma}&=&-3\beta\partial\gamma-2\partial\beta\gamma,
\end{eqnarray}
where $q$ is the background charge of $T_{\phi}$. The first
nilpotency condition $Q_{0}^{2}=0$ requires that the total central
charge vanishes, i.e.,
\begin{equation}
-\frac{517}{60}+\frac{c_{eff}}{12}+q^{2}=0.
\end{equation}

Next, we will construct the explicit form of $Q_{1}$. The most
extensive combinations can be constructed as follows:
\begin{eqnarray}
\aligned
 Q_{1}=\oint dz \gamma\big(&f_{1}W_{L}+f_{2}\partial T_{L}
         +f_{3}T_{L}\partial \phi
         +f_{4}\partial \beta \partial \gamma \\
      &  +f_{5}\partial^{3} \phi
         +f_{6} \partial^{2}\phi \partial \phi
         +f_{7}(\partial \phi)^{3}
         +f_{8} \partial \phi \beta \partial \gamma \big).
\endaligned  \label{W23Q1}
\end{eqnarray}
Then considering the last two conditions in (\ref{nicon}), we
obtain two solutions:

\begin{itemize}

{\item Solution 1}
\begin{equation}
\begin{array}{l} \label{solution1}
  f_{1}=f_{2}=f_{3}=0,\;\;\;
  f_{4}=147 m_{1}, \;\;\;
  f_{5}=38 m_{1} q,\\
  f_{6}=294 m_{1}, \;\;\;
  f_{7}=16 m_{1} q,\;\;\;
  f_{8}=72 m_{1} q,\\
  q^{2}=\frac{49}{8},\;\;\;
  c_{eff}=\frac{299}{10},
\end{array}
\end{equation}
where $m_{1}$ is a non-zero constant. One may note that in this
solution, $Q_1$ has no terms containing $T_{L}$ or $W_{L}$. If
choose $m_{1}=\frac{1}{16 q}$, we will get the analogous results
as Ref \cite{LuNPB1993} for $W_{2,3}$ string. The main difference
is that the ghost fields $(b, c)$ and $(\beta, \gamma)$ here are
introduced for Liouville system.

{\item Solution 2}
\begin{equation}
\begin{array}{l}
  f_{1}=24 \sqrt{5} h m_{2},\;\;\;
  f_{2}=-96 m_{2},\;\;\;
  f_{3}=-30 m_{2} q,\;\;\;
  f_{4}=192 m_{2}, \\
  f_{5}=53 m_{2} q,\;\;\;
  f_{6}=384 m_{2},\;\;\;
  f_{7}=20 m_{2} q,\;\;\;
  f_{8}=90 m_{2} q,\\
  q^{2}=\frac{32}{5},\;\;\;
  c_{eff}=\frac{133}{5},
\end{array}
\end{equation}
where $h^{2}=1$ and $m_{2}$ is a non-zero constant. Different from
the Solution (\ref{solution1}), all coefficients here are
non-vanishing.
\end{itemize}

In Table \ref{Table1}, we give a list of background charges and
central charges of various fields. It is clear that all the
central charges of $T_{eff}$ and $T_{\phi}$ are fractional for
both solutions, which is different from the matter system.

\begin{table}[h]
\caption{Background charges and central charges of various fields
for $W_{2,3}^L$ string. \label{Table1}}
\begin{center}
\renewcommand\arraystretch{1.4}
\begin{tabular}{|c||c|c|}
  \hline
  field & background charge & central charge \\
  \hline
  $T_{L}$ & ~ & $-\frac{22}{5}$ \\
  \hline
  $T_{eff}$(solution 1) & ~ & $\frac{299}{10}$ \\
  $T_{eff}$(solution 2) & ~ & $\frac{133}{5}$ \\
  \hline
  $\phi$(solution 1)  &  $\pm \frac{7}{2\sqrt{2}}$ & $\frac{149}{8}$ \\
  $\phi$(solution 2)  & $\pm 4\sqrt{\frac{2}{5}}$ & $\frac{389}{20}$ \\
  \hline
  $(b, c)$ &  ~ & $-26$ \\
  $(\beta, \gamma)$ & ~ & $-74$ \\
  \hline
\end{tabular}
\end{center}
\end{table}

\subsection{BRST operator for $W_{2,4}^{L}$ string}

From the OPE $W_{L}(z)W_{L}(\omega)$ for $W_{2,4}^L$, one will
find that there are four values of the central charge at which the
$W_{2,4}^{L}$ algebra becomes singular, namely $c_{L}=-24$,
$\frac{1}{2}$, $-\frac{22}{5}$ and $-\frac{68}{7}$. After
rescaling the spin-4 current, it is shown that only at
$c_{L}=-24$, the $W_{2,4}^L$ algebra is consistent and satisfies
the Jacobi identity, this case was found in \cite{LuIPLB1995} and
neglected in \cite{KauschNPB1991,BlumenhagenNPB1991}. Then the OPE
$W_{L}(z)W_{L}(\omega)$ reads
\begin{equation}
W_{L}(z)W_{L}(\omega) \sim \frac{2T_{L} W_{L}-\frac{1}{3}
\partial^{2}W_{L}}{(z-\omega)^{2}}+\frac{\partial (T_{L}W_{L})-\frac{1}{6}
\partial^{3} W_{L}}{z-\omega}. \label{W24W}
\end{equation}
This OPE relation provides us with a way to construct the BRST
operator for the Liouville system of $W_{2,4}^L$ string. Next, we
will give an explicit BRST operator for $W_{2,4}^L$ string under
the grading form (\ref{gradingQB}).

First, $Q_{0}$ takes the form of (\ref{W23Q0}), where $T_{\phi}$
and $T_{bc}$ are the same as $W_{2,3}^L$, while $T_{\beta \gamma}$
is given by
\begin{equation}
T_{\beta\gamma}=-4\beta\partial\gamma-3\partial\beta\gamma.
\end{equation}
Considering the first nilpotency condition $Q_{0}^{2}=0$, we
obtain that the total central charge vanishes, i.e.,
\begin{equation}
-\frac{65}{4}+\frac{c_{eff}}{12}+q^{2}=0.
\end{equation}
This offers us a relation between the central charge $c_{eff}$ of
energy-momentum tensor $T_{eff}$ and the background charge $q$ of
scalar field ${\phi}$. Once obtain the value of $q$, the central
charge $c_{eff}$ can also be obtained through this equation.

Next, we give a mostly extensive combinations of BRST operator
$Q_{1}$ for $W_{2,4}$ string:
\begin{equation} \aligned
Q_{1} = \oint dz \gamma &\big(g_{1} W_{L}
            +g_{2} T_{L}^{2}
            +g_{3} \partial^{2} T_{L}
            +g_{4} \partial T_{L} \partial \phi
            +g_{5} T_{L} \partial^{2} \phi
            +g_{6} T_{L}(\partial \phi)^{2}
            +g_{7} T_{L} \beta \partial \gamma\\
            &+g_{8} \partial^{4} \phi
            +g_{9} (\partial^{2} \phi)^{2}
            +g_{10} (\partial \phi)^{4}
            +g_{11} \partial^{3} \phi \partial \phi
            +g_{12} \partial^{2} \phi (\partial \phi)^{2}
            +fg_{13} (\partial \phi)^{2} \beta \partial \gamma \\
            &+g_{14} \partial^{2} \phi \beta \partial \gamma
            +g_{15} \partial \phi \beta \partial^{2} \gamma
            +g_{16} \partial^{2} \beta \partial \gamma
            +g_{17} \beta \partial^{3} \gamma \big).
\endaligned \end{equation}
Then considering the last two conditions in (\ref{nicon}), we also
obtain two solutions:

\begin{itemize}

{\item Solution 1}
\begin{equation}
\begin{array}{l}
  g_{i}=0\;(i=1-7),\;\;\;
  g_{8}=-779 m_{3}, \;\;\;
  g_{9}=-7590 m_{3}q,\;\;\;
  g_{10}=-900 m_{3} q, \\
  g_{11}=-7020 m_{3} q,\;\;\;
  g_{12}=-43320 m_{3},\;\;\;
  g_{13}=-7200 m_{3} q, \;\;\;
  g_{14}=14820 m_{3},\\
  g_{15}=37620 m_{3},\;\;\;
  g_{16}=-5220 m_{3} q,\;\;\;
  g_{17}=1560 m_{3} q,\\
  q^{2}=\frac{361}{30},\;\;\;
  c_{eff}=\frac{253}{5},
\end{array}
\end{equation}
where $m_{3}$ is a non-zero constant.

{\item Solution 2}
\begin{equation}
\begin{array}{l}
  g_{i}=0\;(i=1-7)\;\;,
  g_{8}=-22356 m_{4}, \;\;\;
  g_{9}=-44280 m_{4}q,\;\;\;
  g_{10}=-54004 m_{4} q, \\
  g_{11}=-44640 m_{4} q,\;\;\;
  g_{12}=-262440 m_{4},\;\;\;
  g_{13}=-43200 m_{4} q, \;\;\;
  g_{14}=116640 m_{4},\\
  g_{15}=233280 m_{4},\;\;\;
  g_{16}=-28800 m_{4} q,\;\;\;
  g_{17}=4320 m_{4} q,\\
  q^{2}=\frac{243}{20},\;\;\;
  c_{eff}=\frac{246}{5},
\end{array}
\end{equation}
where $m_{4}$ is a non-zero constant.
\end{itemize}

Both solutions have $g_{i}=0$ for $i=1-7$. This leads to the
vanishing of $T_{L}$ and $W_{L}$ in $Q_{1}$. One can also see that
these coefficients $g_{i}$ here is more larger and complicated
than the case of $W_{2,3}^L$ string. Choose the exact values for
$m_{3}$ and $m_{4}$, i.e. $m_{3}=-\frac{1}{10830} q$ and
$m_{4}=-\frac{1}{65610} q$, these two solutions is analogous the
results in Ref. \cite{Lu2IJMPA1994} for $W_{2,4}$ strings. The
background charges and central charges for these fields can be
found in Table \ref{Table2}. In \cite{LuIPLB1995}, the ghost
fields $(b, c)$ together with $(\beta, \gamma)$ were used to
construct the BRST operator, and the result for this
$W_{2,4}^{M}\otimes W_{2,4}^{L}$ at central charge $c_{L}=-24$ is
given by
\begin{equation}
 Q=Q_{0}+Q_{1}-\oint dz \gamma \bigg(\frac{167}{22} W_{L}
   +\frac{27889}{484}W_{L} b \partial \gamma\bigg),
\end{equation}
where the first two terms are for the matter system.

\begin{table}[h]
\caption{Background charges and central charges of various fields
for $W_{2,4}^L$ string. \label{Table2}}
\begin{center}
\renewcommand\arraystretch{1.4}
\begin{tabular}{|c||c|c|}
  \hline
  field &  background charge & central charge \\
  \hline
  $T_{L}$ & ~ & $-24$ \\
  \hline
  $T_{eff}$(solution 1) & ~ & $\frac{253}{5}$ \\
  $T_{eff}$(solution 2) & ~ & $\frac{246}{5}$ \\
  \hline
  $\phi$(solution 1)  & $\pm \frac{19}{\sqrt{30}}$ & $\frac{727}{10}$ \\
  $\phi$(solution 2)  & $\pm \frac{9\sqrt{3}}{2\sqrt{5}}$ & $\frac{367}{5}$ \\
  \hline
  $b, c$ & ~ & $-26$ \\
  $\beta, \gamma$ & ~ & $-146$ \\
  \hline
\end{tabular}
\end{center}
\end{table}

\subsection{BRST operator for $W_{2,6}^{L}$ string}

It is worth to point that $W_{2,5}^L$ algebra does not exist at
the quantum level for the value of central charge required by the
criticality. But one can obtain the nilpotent quantum BRST
operator by adding $\hbar$-dependent corrections for it. The
explicit result was given in \cite{LuIPLB1995} and we will not
discuss it in detail.

Next, we consider the case of $W_{2,6}^L$. We expect that it has
the same graded form. At $c_{L}=-2$ and $c_{L}=-\frac{286}{3}$,
the $W_{2,6}^L$ algebra becomes degenerate. At $c_{L}=-2$, the OPE
of the spin-6 current with itself is of the following form
\cite{LuIPLB1995}:
\begin{equation}
W_{L}(z)W_{L}(\omega) \sim \frac{2
\Lambda}{(z-\omega)^{2}}+\frac{\partial \Lambda}{z-\omega},
\label{W261}
\end{equation}
where $\Lambda =T_{L}^{2}W_{L}-\frac{5}{9}T_{L}
\partial^{2}W_{L}+\frac{9}{19}\partial T_{L} \partial W_{L}-\frac{17}{6}\partial^{2}T_{L} W_{L}
+\frac{1}{36} \partial^{4} W_{L}$. At $c_{L}=-\frac{286}{3}$, the
OPE is given by
\begin{equation}\aligned
 W_{L}(z)W_{L}(\omega) &\sim \frac{W_{L}}{(z-\omega)^{6}}+\frac{1}{2}
                     \frac{\partial W_{L}}{(z-\omega)^{5}}+\frac{7}{52} \frac{\partial^{2}
                     W_{L}}{(z-\omega)^{4}}+\frac{1}{39} \frac{\partial^{3}
                     W_{L}}{(z-\omega)^{3}}+\frac{1}{260} \frac{\partial^{4}
                     W_{L}}{(z-\omega)^{2}}\\
                &+\frac{1}{2080} \frac{\partial^{5} W_{L}}{z-\omega} -\frac{9}{35}
                     \frac{\Lambda_{1}}{(z-\omega)^{4}}-\frac{9}{70} \frac{\partial
                     \Lambda_{1}}{(z-\omega)^{3}}-\frac{81}{2380} \frac{\partial^{2}
                     \Lambda_{1}}{(z-\omega)^{2}}-\frac{3}{476} \frac{\partial^{3}
                     \Lambda_{1}}{z-\omega}\\
                &+\frac{27}{700} \frac{\Lambda_{2}}{(z-\omega)^{2}}+\frac{27}{1400}
                     \frac{\partial \Lambda_{2}}{z-\omega}, \label{W262} \endaligned
\end{equation}
where $\Lambda_{1}$ and $\Lambda_{2}$ have spins 8 and 10
respectively, they are given by
\begin{eqnarray}
\Lambda_{1}&=& T_{L} W_{L}-\frac{3}{26}\partial^{2} W_{L}, \\
\Lambda_{2}&=&T_{L}^{2}W_{L}-\frac{35}{153}T_{L}\partial^{2}W_{L}-\frac{2}{153}
\partial T_{L} \partial W_{L}-\frac{29}{102} \partial^{2}T_{L} W_{L}+\frac{7}{612}
\partial^{4} W_{L}.
\end{eqnarray}
It is clear that every term on the right hand side of the OPEs
(\ref{W261}) and (\ref{W262}) has $W_L$, so we can consistently
set it to zero, and the BRST operators will continue to be
nilpotent.

Now, using the OPEs relations (\ref{W261}) and (\ref{W262}), we
would like to construct the BRST operator of Liouville system for
$W_{2,6}$ string at $c_{L}=-2$ and $c_{L}=-\frac{286}{3}$. The
BRST operator are given in the grading form
\begin{eqnarray}
 Q_{B}&=& Q_{0}+Q_{1}, \\
 Q_{0}&=&\oint dz c(T_{eff}+T_{L}-6\beta \partial \gamma
         -5\partial\beta \gamma-b\partial c). \\
 Q_{1}&=&\oint dz \gamma \big(h_{1} W_{L}
     +h_{2}T_{L}^{3}
     +h_{3} \partial^{4} T_{L}
     +h_{4} \partial^{2} T_{L} T_{L}
     +h_{5} (\partial T_{L})^{2}
     +h_{6} T_{L} \partial^{2} \beta \partial\gamma \nonumber \\
    &&+h_{7} T_{L} \beta \partial^{3} \gamma
     +h_{8} T_{L} \partial \beta \partial^{2} \gamma
     +h_{9} T_{L}^{2} \beta \partial \gamma
     +h_{10} \partial^{3} \beta \partial^{2}\gamma
     +h_{11} \partial^{2} \beta \partial^{3} \gamma \nonumber \\
    &&+h_{12} \partial \beta \partial^{4} \gamma
     +h_{13} \beta \partial^{5} \gamma
     +h_{14} \partial \beta \beta
             \partial^{2}\gamma\partial\gamma \big).
\end{eqnarray}
Considering the nilpotency condition $Q_{0}^{2}=0$, we get
\begin{equation}
c_{L}+c_{eff}=388.
\end{equation}
This implies that $c_{eff}=390$ and $\frac{1450}{3}$ at $c_{L}=-2$
and $-\frac{286}{3}$, respectively. Using the nilpotency
conditions $Q_{1}^{2}=0$ and $\{Q_{0},Q_{1}\}=0$, we could
determine the coefficients $h_i$. For both cases of $c_{L}=-2$ and
$-\frac{286}{3}$, these coefficients are
\begin{equation}
h_{i}=0\;(i=1-9,13,14),\;\;\;
h_{10}=m_{5},\;\;\;h_{11}=2m_{5},\;\;\;h_{12}=m_{5},
\end{equation}
where $m_{5}$ is a non-zero constant.

Then $Q_{1}$ is given by
\begin{equation}
 Q_{1}=\oint dz \gamma (m_{5} \partial^{3} \beta \partial^{2}\gamma
     +2m_{5} \partial^{2} \beta \partial^{3}
     \gamma+m_{5} \partial \beta \partial^{4} \gamma).
\end{equation}
This result shows that $W_{L}$ and $T_{L}$ don't appear in the
final expression of $Q_{1}$. One of the main reasons may be that
more strict conditions are required in the grading form.

\section{Conclusion}
\label{secConclusion}

In this paper, we mainly investigate the $W_{2,s}^{M}\otimes
W_{2,s}^{L}$ system in which these two sub-systems generate two
different $W_{2,s}$ algebras. We first give a brief discussion on
the physical states of $W_{2,s}$ strings. These physical states
are characterized by level number $\ell$ and ghost number $G$. The
ground state is denoted by $\ell=0$, and $G=0$. The higher states
can be obtained by acting on the ground $|\downarrow,k\rangle$
with descendent operators. Using the condition
$Q_{B}|\psi\rangle=0$ and the non-trivial condition $|\psi\rangle
\neq Q_{B}|\varphi\rangle$, one can get the explicit form of
higher level states. The physical states up to level $\ell= 2$ are
given in (\ref{state}), where the spin-2 and spin-3 currents are
`frozen'.

Then the realizations of non-linear quantum $W_{2,s}$ algebras
from linear $W_{1,2,s}$ algebras are discussed. The BRST operators
of $W_{2,s}$ algebras are also obtained through the realizations
of the linear $W_{1,2,s}$ algebras. For the case of $s=3$ and 4,
the realizations of these linear algebras are got by introducing
ghost-like fields. In fact, this is not the only way to construct
the linear $W_{1,2,s}$ algebras. In Refs.
\cite{LiuJHEP2005,ZhangCTP2006}, by introducing spinor fields, we
gave the explicit realizations of linear $W_{1,2,s}$ algebras.
When $s=5$, there exists no such linear algebra. Fortunately, the
linear $W_{1,2,6}$ algebra exists, but the central charge takes an
exact value, i.e., $c=390$. By introducing a pair of bosonic
ghost-like fields $(R, S)$ with spins $(6,-5)$ and a pair of
fermionic ghost-like fields $(b_{1},c_{1})$ with spins $(k,1-k)$,
we obtain a new realization of $W_{1,2,6}$ algebra. A new BRST
operator for $W_{2,6}$ algebra can be given out by making an
invert of the basis.

We also obtain the BRST operators for the Liouville system of
$W_{2,s}^{M}\otimes W_{2,s}^{L}$. We find that the coefficients
appeared in the $W_{2,3}^L$ algebra will become divergent and the
algebras will be singular at $c_{L}=-\frac{22}{5}$. But one can
rescale the spin-3 current such that its OPE with itself has no
divergent coefficients. This provides us a way to construct the
$W_{2,3}^{L}$ string at $c_{L}=-\frac{22}{5}$, and we obtained two
solutions for this case. For the $W_{2,4}^L$ algebra, the
singularity appears at $c_{L}=-24$, $\frac{1}{2}$,
$-\frac{22}{5}$, and $-\frac{68}{7}$. After rescaling the spin-4
current, the algebra will satisfy the Jacobi identity only at
$c_{L}=-24$, and the two explicit forms of BRST operators for
$W_{2,4}^{L}$ string at $c_{L}=-24$ are given. At last, we
construct the BRST operator for $W_{2,6}^{L}$ at $c_{L}=-2$ and
$c_{L}=-\frac{286}{3}$. It is worth to point that in all these
constructions, the BRST grading method is used.

\section*{Acknowledgements}

This work was supported by the National Natural Science Foundation
of China (No. 10705013), the Doctor Education Fund of Educational
Department of China (No. 20070730055) and the Fundamental Research
Fund for Physics and Mathematics of Lanzhou University (No.
Lzu07002). L.J. Zhang acknowledges financial support from Shanghai
Education Commission.

\end{document}